\documentclass[lettersize,journal]{IEEEtran}
\usepackage{amsmath,amsfonts}
\usepackage{algorithmic}
\usepackage{algorithm}
\usepackage{array}
\usepackage{multirow}
\usepackage{makecell}
\usepackage[caption=false,font=footnotesize,labelfont=sf,textfont=sf]{subfig}
\usepackage{textcomp}
\usepackage{stfloats}
\usepackage{tabularx,booktabs} %Self added 
\usepackage{verbatim}
\usepackage{graphicx}
\usepackage{cite}
\usepackage{url}
\usepackage{relsize} %Self added
\usepackage{color}
\usepackage{tabularray}
\usepackage{newtxtext} % Provides Times font with small caps italic
\makeatletter
\let\NAT@parse\undefined
\makeatother
\usepackage{hyperref}  %hyperref still needs to be put at the end!

\definecolor{b}{rgb}{0,0,0} %%Pure Version

\begin{document}

\title{Terra-Fi: Decoupling Sensing and Wireless Depths for Sustainable Deep Underground Monitoring}

%Terra-Fi: Decoupling Sensing and Wireless Depths for Sustainable Deep Underground Monitoring

%Terra-Fi: A Shallow-Hub and Deep-Sensing Architecture for Sustainable Underground Monitoring

\author{Kaiqiang Lin %\IEEEmembership{Student~Member,~IEEE,}
        and Mohamed-Slim Alouini%\IEEEmembership{Student~Member,~IEEE,}
        
\thanks{This work was supported in part by King Abdullah University of Science and Technology. (\textit{Corresponding author: Kaiqiang Lin})}      

\thanks{Kaiqiang Lin and Mohamed-Slim Alouini are with King Abdullah University of Science and Technology, Kingdom of Saudi Arabia.}
}


% column for its text to clear the IEEEpubid mark
\maketitle

\begin{abstract}
Wireless underground sensor networks (WUSNs) enable in-situ monitoring of subsurface processes and buried assets to deliver substantial social and economic benefits across diverse verticals. However, meter-scale underground monitoring remains challenging since conventional WUSNs place the entire underground device at the target sensing depth, where severe soil-induced attenuation significantly degrades communication reliability and makes wireless energy transfer (WET) inefficient and even impractical. To address this limitation, we propose \emph{Terra-Fi}, a novel shallow-hub and deep-sensing architecture that decouples the sensing depth from the wireless communication and energy-replenishment depth. In \emph{Terra-Fi}, a shallow wireless hub is deployed at a depth favorable for communication, WET, and maintenance, while application-specific sensing probes are extended to meter-scale target layers through power-and-signal feeders. On this basis, we first examine the depth limitations of underground communication and WET and compare representative monitoring architectures. Then, we present its system architecture, operation flow, and potential benefits. Through a precision agriculture case study, our numerical results reveal that \emph{Terra-Fi} establishes more robust wireless connectivity and a longer operational lifetime than conventional fully buried WUSN architectures, particularly at greater sensing depths and higher volumetric water content. Moreover, its performance gain can be enhanced through time allocation and transmission-configuration optimization. Finally, we pinpoint practical challenges and future research directions for deploying \emph{Terra-Fi} into WUSN systems.
\end{abstract}
%\begin{IEEEkeywords}Article submission, IEEE, IEEEtran, journal, \LaTeX, paper, template, typesetting.
%\end{IEEEkeywords}
\section{Introduction}
Wireless underground sensor networks (WUSNs) are opening a new sensing frontier beneath the ground surface, where critical physical processes and buried assets remain largely hidden from conventional remote sensing technologies~\cite{LinWCM}. By enabling underground devices (UDs) to collect subsurface information and wirelessly deliver it to aboveground gateways or base stations, WUSNs facilitate applications such as precision agriculture, underground infrastructure monitoring, and post-disaster geohazard warning~\cite{LinMag}. Unlike terrestrial monitoring, subterranean sensing is inherently depth-dependent as the target process often occurs in an application-specific subsurface layer, which can generally reach meter-scale depths~\cite{WUSNReview}. Specifically, such depth requirements arise in the following representative scenarios:
\begin{itemize}
    \item \textbf{Precision Agriculture:} Precision agriculture depends on root-zone measurements of soil moisture, temperature, salinity, and nutrient status to quantify crop water uptake and growth stress. These variables are distributed within the active rooting region, which can extend to $1$--$1.8$~m depending on crop type~\cite{CroprootDepth}.

    \item \textbf{Underground Pipeline Monitoring:} Pipeline leakage detection relies on pressure/flow anomalies, pipe-wall acoustic signatures, and moisture or gas variations in the surrounding soil. Since these indicators originate from the buried pipe and its adjacent soil environment, they are commonly associated with depths of $0.6$--$1.8$~m below the ground surface~\cite{PipeDepth}.

    % \item \textbf{Environmental Surveillance:} Environmental surveillance requires depth-resolved measurements of soil temperature, volumetric water content (VWC), salinity, and related hydrological states to characterize subsurface heat--water transport, salinity evolution, and shallow environmental anomalies. In national-scale soil observatories, these variables are monitored along vertical soil profiles down to approximately $2$~m~\cite{SoilDepth}.

    \item \textbf{Geohazard Early Warning:} Rainfall-induced shallow landslides are governed by variations in matric suction, pore-water pressure, soil temperature, VWC, and deformation within the potential sliding zone. These precursors typically occur at depths of approximately $1$--$3$~m. For instance, the field study in~\cite{LandDepth} deployed suction sensors at depths of $0.95$--$2.55$~m and inferred a sliding-zone depth of $1.5$~m.
\end{itemize}

However, most existing WUSN designs rely on a monolithic architecture, in which sensing, communication, energy storage, and power replenishment are co-located within a single device deployed at the target sensing depth~\cite{VuranWUSNsreview}. This tight functional integration becomes a fundamental bottleneck for meter-scale underground sensing, as severe soil-induced attenuation substantially weakens underground-to-aboveground (UG2AG) wireless links and drastically reduces the efficiency of wireless energy transfer (WET). Although higher transmit power, higher-gain antennas, and multi-antenna techniques can partially compensate for these propagation losses, they inevitably increase energy consumption, hardware complexity, and deployment cost.

To overcome this depth mismatch, we propose \emph{Terra-Fi}, a shallow-hub and deep-sensing architecture in which communication, control, energy storage, and WET functions are retained at a shallow depth, while sensing probes are extended to deeper target layers through protected power-and-signal feeders. By separating sensing depth from the wireless and energy-service depth, \emph{Terra-Fi} enables meter-scale in-situ monitoring without forcing the complete UD to operate under the most severe underground conditions. To the best of the authors' knowledge, this is the first work to propose \emph{Terra-Fi} as a depth-decoupled WUSN architecture for sustainable deep underground monitoring and to validate its advantages over conventional fully buried WUSN architectures. The main contributions of this article are summarized as follows:
\begin{enumerate}
    \item We discuss the depth limitations of underground wireless communication and WET and identify the sensing-service depth mismatch in conventional underground IoT architectures.
    
    \item We conceptualize the \emph{Terra-Fi} architecture that decouples deep in-situ sensing from shallow wireless communication and energy replenishment. Moreover, we present its system architecture, operation flow, and key advantages relative to existing underground monitoring architectures.

    \item  We develop a simulation framework for center-pivot irrigation farming scenarios and quantitatively compare \emph{Terra-Fi} with conventional fully buried WUSNs in terms of communication reliability and operational lifetime under different underground conditions and network scales. We further pinpoint open challenges and future research directions toward practical \emph{Terra-Fi} deployment.
\end{enumerate}

\section{Depth Limits of Underground Wireless Operation}
This section examines the depth limitations of underground wireless communication and WET, and compares representative underground monitoring architectures to motivate the proposed \emph{Terra-Fi} design.

\subsection{Depth-Limited Wireless Communication}
% WUSNs generally adopt two networking approaches: ad-hoc networking and centralized networking~\cite{LinMag}. In ad-hoc WUSNs, UDs relay data through underground multi-hop links; however, the practical communication distance between UDs is typically limited to a few meters, which makes multi-hop underground networking difficult to scale for wide-area monitoring. In contrast, centralized WUSNs allow UDs to transmit data directly to aboveground gateways or base stations, thereby avoiding power-intensive underground relaying. Therefore, centralized low-power wide-area network (LPWAN) connectivity is often preferred for reliable and energy-efficient underground monitoring.
WUSNs generally adopt either underground multi-hop or centralized networking approaches~\cite{LinMag}. Multi-hop schemes extend coverage through buried relays but incur additional underground communication and energy costs, whereas centralized WUSNs directly connect UDs to aboveground gateways or base stations. Consequently, centralized low-power wide-area network (LPWAN) connectivity is often preferred for reliable and energy-efficient underground monitoring.

Among LPWAN technologies, LoRaWAN and NB-IoT are representative candidates for UG2AG communication. LoRaWAN supports flexible private deployment, whereas NB-IoT provides operator-supported wide-area connectivity but depends on existing network availability. However, both are strongly constrained by burial depth and soil conditions. At $433$~MHz, LoRaWAN achieved only a $16$~m UG2AG range at a burial depth of $0.8$~m and $15\%$ VWC~\cite{LinLoRaWUSNs}. For NB-IoT at $716$~MHz, the reachable burial depth decreased from $0.7$~m at $5\%$ VWC to $0.3$~m at $25\%$ VWC~\cite{NB-IoTDepth}. These results indicate that deploying fully buried WUSNs at meter-scale depths would substantially exacerbate underground propagation loss, making reliable and scalable underground monitoring increasingly impractical. In subterranean wireless channels, soil-induced attenuation increases rapidly with burial depth and is further governed by VWC, soil texture, operating frequency, and antenna--soil coupling. Although higher transmit power, higher-gain antennas, or denser gateway deployments can improve the link budget, these link-level remedies increase energy consumption, hardware complexity, and infrastructure cost. This depth-dependent limitation motivates separating the sensing depth from the wireless-operation depth, with probes placed at the target layer and radio components retained in shallower soil.

\subsection{Depth-Limited Wireless Energy Transfer}
Severe soil-induced attenuation increases the energy required for reliable communication, while limited battery capacity and impractical excavation-based battery replacement make long-term operation of UDs particularly challenging at meter-scale depths. To extend the lifetime of UDs, various energy-supply techniques have been investigated for underground scenarios, which can be broadly classified into ambient energy harvesting and dedicated external energy replenishment~\cite{SalamWET}. Although ambient energy harvesting has been explored, its available power is generally weak and intermittent~\cite{SalamWET}. Dedicated WET therefore provides a more controllable replenishment mechanism, with magnetic-induction WET (MI-WET) and radio-frequency WET (RF-WET) representing two major approaches.

MI-WET exploits near-field magnetic coupling and can achieve high power-transfer efficiency over short distances but requires close coil separation and accurate alignment. For instance, a $13.56$~MHz MI-WET prototype delivered a peak received power of $30$~W to a UD buried at $0.2$~m depth with a coil separation of $0.25$~m~\cite{MIUWET}. By contrast, RF-WET offers greater deployment flexibility via fixed power beacons or UAV-mounted chargers, but it is severely constrained by soil absorption. The UAV-assisted underground RF-WET field experiment in~\cite{UAVRFUWET} demonstrated that UDs buried at depths of $0.15$~m and $0.30$~m with a VWC of $40\%$ received average RF powers of $2$~dBm and $-5$~dBm, respectively, under a $915$~MHz power source operating at a transmit power of $36$~dBm.

The aforementioned WET solutions are feasible for shallow underground energy replenishment, typically within several tens of centimeters, but become inefficient as UDs approach meter-scale depths. RF-WET is constrained by harvester sensitivity, which is generally more stringent than communication receiver sensitivity, whereas MI-WET is limited by near-field coupling, coil separation, and alignment accuracy. Increasing the transmit power of energy sources, antenna/coil aperture, charger density, or charging duration can improve energy delivery, but only with higher energy consumption, hardware complexity, and deployment cost, without removing soil-induced WET loss. This calls for a shallow-energy and deep-sensing architecture, in which energy harvesting, storage, and power-management modules are placed in a shallow hub with favorable WET conditions, while power-and-signal feeders deliver energy and control to sensing probes deployed at meter-scale target depths. This design also enables easier inspection and battery replacement without full-depth excavation.
% Existing wireless communication and WET technologies are effective at shallow burial depths, whereas many underground monitoring applications require sensing access at meter-scale depths. This discrepancy creates a depth mismatch in conventional fully buried WUSNs. Specifically, the sensing probe must reach the target layer, while the communication, energy harvester, and power-management modules are more reliable, efficient, and serviceable when placed closer to the surface. This observation motivates the \emph{Terra-Fi} platform to decouple deep sensing from shallow wireless communication and energy-service functions.

\subsection{Existing Underground Monitoring Architectures}
Underground monitoring systems generally adopt either cabled or fully buried wireless architectures, as summarized in Table~\ref{tab:architecture_comparison}. Conventional cabled systems connect buried probes to aboveground loggers, avoiding underground wireless propagation and facilitating access to the logger and power supply. However, exposed surface electronics sacrifice concealment, remain vulnerable to agricultural machinery, weather, theft, and vandalism, and may interfere with field operations~\cite{IFAWUSNReview}. Fully buried WUSNs address these concerns by placing sensing, communication, antennas, and energy storage underground, thereby preserving an unobstructed and concealed surface. This forces the complete UD to operate at the sensing depth, where wireless communication, WET, and maintenance become constrained. Multi-hop variants shorten individual underground links through buried relays, but require additional radios, energy sources, routing, and distributed maintenance~\cite{Cotrim2024}.

\emph{Terra-Fi} occupies an intermediate position between surface-connected sensing systems and fully buried WUSNs. It retains communication, control, energy storage, and WET in a shallow buried hub, while deeper probes are connected through protected power-and-signal feeders. This preserves surface concealment, reduces underground propagation loss, and avoids additional buried relays, but introduces feeder integration, sealing, and installation requirements.

% \emph{Terra-Fi} occupies an intermediate architectural regime between surface-connected sensing systems and fully buried WUSNs. By relocating communication, control, energy storage, and WET subsystems to a shallow buried hub, it preserves surface concealment while reducing the severe underground propagation. Compared with multi-hop WUSNs, \emph{Terra-Fi} also eliminates the need for additional buried relays and the associated routing and energy-management overhead. Deep sensing is instead realized through protected power-and-signal feeders that connect one or more meter-scale probes to the shallow hub. This functional partitioning enables deep sensing while retaining wireless access, energy replenishment, and maintenance at a shallow depth, at the cost of additional feeder integration, sealing, and installation requirements.

\begin{table*}[t]
\centering
\caption{Comparison of representative underground monitoring architectures.}
\label{tab:architecture_comparison}
% \scriptsize
\setlength{\tabcolsep}{1.2pt}
\renewcommand{\arraystretch}{1.05}
\begin{tabular}{>{\centering\arraybackslash}m{2.15cm}>{\centering\arraybackslash}m{1.4cm}>{\centering\arraybackslash}m{1.55cm}>{\centering\arraybackslash}m{1.75cm}>{\centering\arraybackslash}m{1.85cm}>{\centering\arraybackslash}m{1.55cm}>{\centering\arraybackslash}m{1.70cm}>{\centering\arraybackslash}m{2.65cm}>{\centering\arraybackslash}m{2.70cm}}
\hline
\textbf{Architecture} & \textbf{Radio location} & \textbf{Surface electronics} & \textbf{Deep active electronics} & \textbf{Multi-depth sensing} & \textbf{Wireless backhaul} & \textbf{Wireless charging} & \textbf{Key advantage} & \textbf{Key limitation} \\
\hline
Surface logger with cabled probes~\cite{IFAWUSNReview} & Aboveground & Required & Sensing and interface electronics & Supported by multiple probes & Optional at logger & Generally unnecessary & Avoids underground attenuation and provides direct access to the logger and power supply & Limited concealment and vulnerable to machinery, weather, theft, and vandalism \\

Fully buried WUSN~\cite{VuranWUSNsreview} & Sensing depth & Not required & UD & Typically requires separate UDs & Direct UG2AG link & Strongly depth-limited & Preserves surface concealment without wired connections & Communication, WET, and maintenance become difficult with burial depth \\

Fully buried multi-hop WUSN~\cite{Cotrim2024} & Sensing and relay depths & Not required & UD with relay functions & Supported through multiple UDs & Underground relaying to UG2AG & Strongly depth-limited & Concealment while shortening wireless links & Additional buried radios, energy sources, and routing \\

\textit{Terra-Fi} & Shallow buried hub ($\sim$0.3~m) & Not required & Sensing probe and local interface & Multiple probes share one hub & Direct UG2AG link from hub & More favorable at shallow hub &  Preserves concealment while enabling meter-scale sensing with accessible maintenance & Additional feeder integration, sealing, and installation requirements\\
\hline
\end{tabular}
\end{table*}

\section{\emph{Terra-Fi} Platform: A Shallow-Hub and Deep-Sensing Architecture}
\subsection{System Architecture}
\begin{figure*}[!t]
\centering
\includegraphics[width=6in]{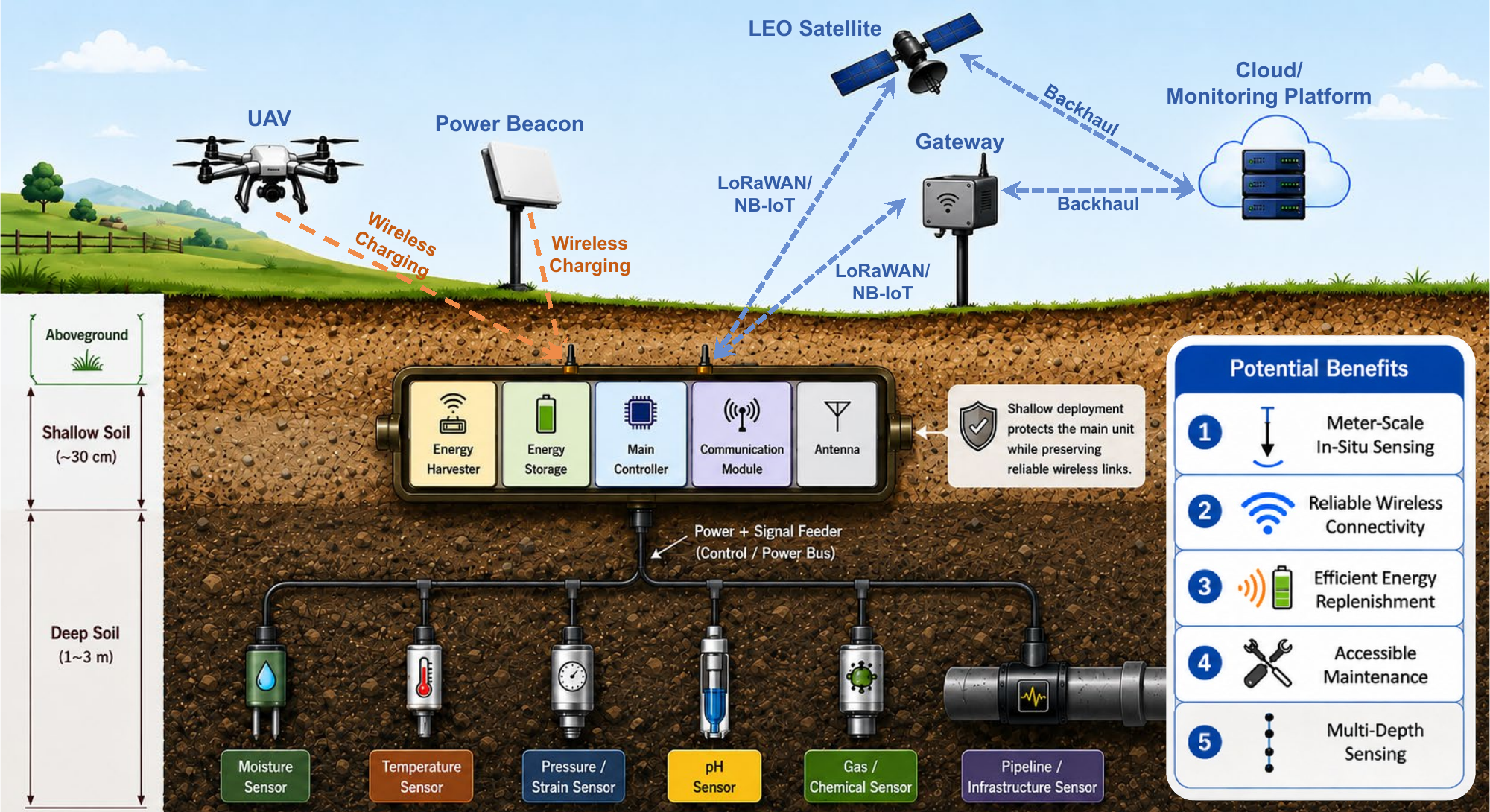}
\caption{Proposed \emph{Terra-Fi} architecture for meter-scale underground monitoring, where a shallow buried hub is connected to deeper sensing probes through protected power-and-signal feeders.}
\label{fig_arc}
\end{figure*}

The key principle of \emph{Terra-Fi} is functional depth decoupling. Application-specific sensing probes remain at the required target depths, whereas communication, control, energy storage, and wireless energy-replenishment functions are consolidated in a shallow buried hub. As illustrated in Fig.~\ref{fig_arc}, the architecture consists of an aboveground service segment, a shallow hub, a power-and-signal feeder and a deep sensing segment.

\begin{itemize}
    \item \textbf{Aboveground Service Segment:} This segment provides external energy, wireless connectivity, and data-processing services to the shallow hub. Energy replenishment is enabled by fixed RF power beacons and UAV-mounted energy transmitters. Fixed beacons support periodic or continuous WET in infrastructure-supported areas, whereas UAV-mounted chargers provide on-demand replenishment for remote or hard-to-access deployments. Terrestrial gateways or base stations connect the shallow hub through LPWAN technologies such as LoRaWAN and NB-IoT, while LEO satellites can extend coverage where terrestrial infrastructure is unavailable or impractical. The collected data are then forwarded through backhaul links to a cloud-based monitoring platform for storage, decision-making, and remote control.
%Fixed beacons utilize RF-WET or MI-WET technologies to periodically or continuously charge nearby shallow hubs in infrastructure-supported areas, whereas UAV-mounted chargers provide on-demand charging for remote, temporary, or hard-to-access deployments. 

    \item \textbf{Shallow Wireless Hub:} This segment serves as the energy, control, and communication center of the \emph{Terra-Fi} platform. In this work, the shallow hub is placed at a representative depth of $0.3$~m, which preserves surface concealment while reducing the underground propagation path for wireless communication and WET and facilitating maintenance. The hub integrates the communication module and antenna, controller, energy storage, power-management circuitry, WET receiver, and feeder interfaces within a sealed, corrosion-resistant enclosure. It harvests and stores energy, coordinates sensing and data transmission, and provides regulated power and signal conditioning for the power-and-signal feeders linking the hub to deep sensing probes.
    % It harvests energy from fixed power beacons or mobile chargers, manages the operation of deep sensing probes, aggregates the sensing data, and wirelessly delivery the data to aboveground receivers.

    \item \textbf{Power-and-Signal Feeder:} This segment provides the power-and-signal connection between the shallow hub and deep sensing probes. The feeder delivers regulated power and bidirectional low-rate data, while its electrical design is governed by conductor resistance, probe power demand, and signal-integrity requirements. To support heterogeneous sensors, probe-side interfaces provide voltage regulation, signal conditioning, and protocol conversion. For long-term underground operation, the feeder assembly must also incorporate sealing, corrosion protection, and mechanical reinforcement against moisture ingress, soil compaction, and installation-induced stress.
    
    \item \textbf{Deep Sensing Segment:} This segment provides direct in-situ access to application-specific target depths through sensing probes for variables such as VWC, temperature, strain, pH, and gas concentration. Only the sensing element and necessary local interface remain at depth, while communication, energy storage, control, and WET functions are retained in the shallow hub. Multiple probes can be deployed at different depths and share the same hub, enabling vertical subsurface profiling without requiring a complete UD at each sensing layer.
\end{itemize}

\subsection{Operation Flow}
The operation of \emph{Terra-Fi} follows an energy-aware sensing and communication cycle.
\begin{enumerate}
    \item The shallow hub harvests energy from fixed or mobile chargers and stores it for sensing, control, and data transmission. According to the available energy and application requirements, the hub activates the required probes and supplies them with regulated power and control commands through the power-and-signal feeders.

    \item The probes acquire in-situ measurements at their target depths and return the sensing data to the shallow hub through the feeders. The hub aggregates and preprocesses the data before transmitting them to terrestrial gateways, base stations, or LEO satellites, depending on network availability.

    \item The aboveground receivers forward the collected data to a cloud-based monitoring platform for storage and intelligent analytics. By integrating AI-enabled analytics with digital-twin models, the platform can support condition assessment, predictive warning, and remote decision-making. Furthermore, it can send derived control commands back to the shallow hub to adapt the reporting period, transmission configuration, and charging schedule according to the monitored state and energy availability, thereby prolonging the operational lifetime of UDs.
\end{enumerate}

\subsection{Architectural Benefits and Tradeoffs}
By separating sensing depth from communication, energy-replenishment, and maintenance functions, \emph{Terra-Fi} combines several advantages of cabled and fully buried monitoring architectures while avoiding some of their major limitations.

\begin{itemize}
    \item \textbf{Meter-Scale In-Situ Sensing:} Similar to cabled probe systems, \emph{Terra-Fi} extends sensing access to meter-scale target depths, but retains the hub below ground to preserve surface concealment. Unlike fully buried WUSNs, only the sensing probes and local interfaces operate at the deepest layers, while communication, control, and energy functions remain at shallow depth.

    \item \textbf{Improved Wireless Connectivity:} Fully buried WUSNs place radios and antennas at the sensing depth, where soil attenuation increases rapidly with depth and VWC. \emph{Terra-Fi} instead retains these components in the shallow hub, reducing the underground propagation length without relying on additional buried relays, higher transmit power, or denser gateways.

    \item \textbf{Improved Energy Replenishment:} In fully buried WUSNs, wireless energy replenishment at meter-scale depths is fundamentally limited by soil-induced attenuation and alignment sensitivity. \emph{Terra-Fi} mitigates these constraints by relocating the WET receiver, energy-storage unit, and power-management circuitry to the shallow hub, while electrical power is delivered to the deeper sensing probes through the feeders.

    \item \textbf{Accessible Maintenance:} Surface-logger systems provide direct access to electronics but sacrifice concealment, whereas fully buried WUSNs may require full-depth excavation for servicing. \emph{Terra-Fi} provides an intermediate solution by keeping serviceable communication and energy components in the shallow buried hub, thereby reducing routine maintenance depth while preserving an unobstructed surface.

    \item \textbf{Shared Multi-Depth Sensing:} Fully buried WUSNs typically require separate UDs at different sensing depths. In \emph{Terra-Fi}, multiple depth-distributed probes can share the communication, control, and energy resources of a single shallow hub, enabling vertical subsurface profiling with fewer UDs.
\end{itemize}

\section{Proof-of-Concept Case Study: Precision Agriculture}

\subsection{Simulation Scenarios and Configurations}
\begin{figure}[!t]
\centering
\includegraphics[width=3.45in]{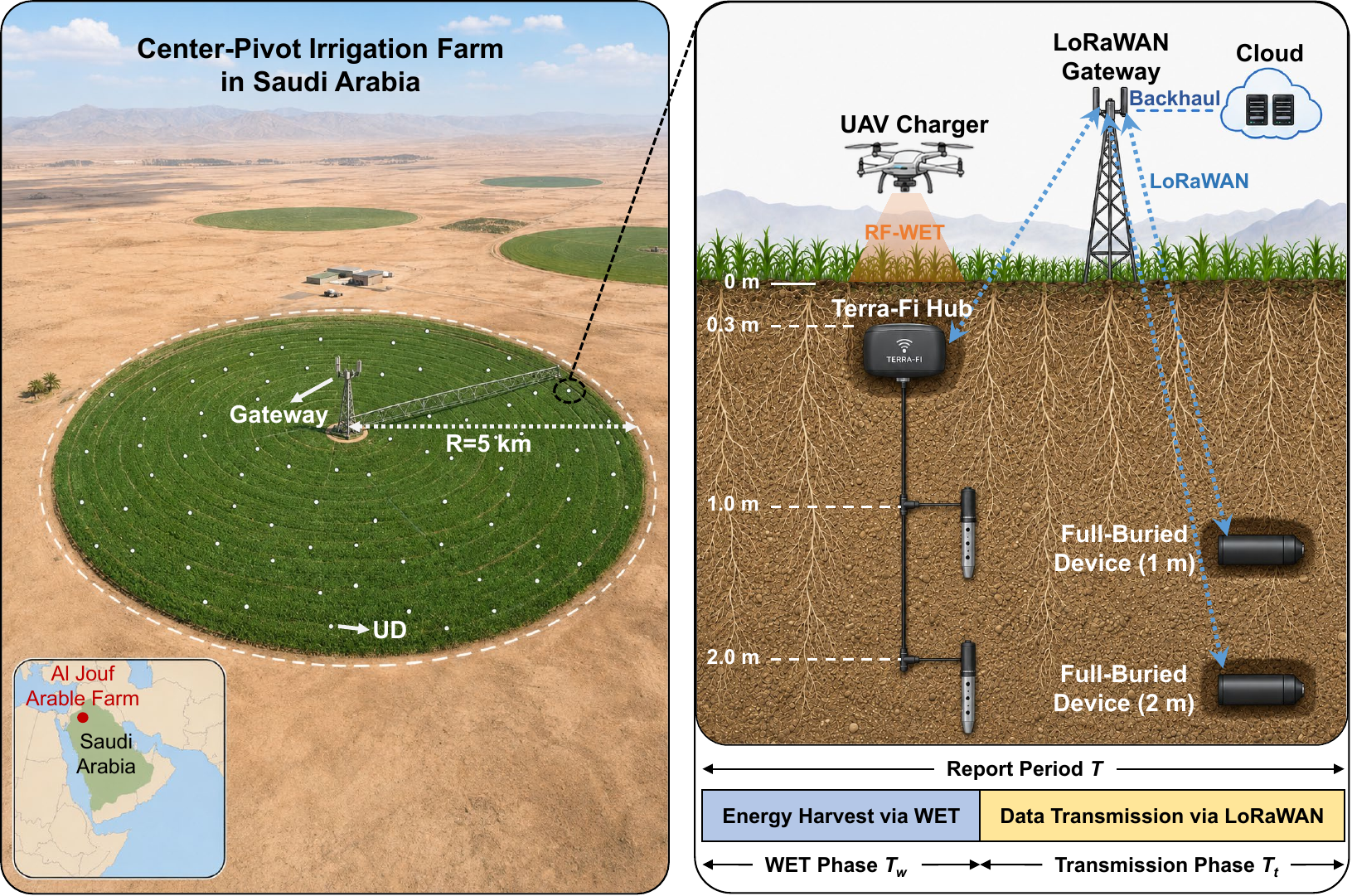}
\caption{Simulated center-pivot irrigation farm scenario in Saudi Arabia and the corresponding system model, where UDs are deployed at the target depth in fully buried WUSNs, whereas \emph{Terra-Fi} places a shallow hub at a depth of $0.3$~m to harvest energy from UAVs via RF-WET during $T_w$ and uses the harvested energy to sense and transmit a $10$-byte packet to the LoRaWAN gateway during $T_t$ within a reporting period of $T$.}
\label{fig_simscenarios}
\end{figure}

To gain practical insight into the feasibility of \emph{Terra-Fi} for long-term underground monitoring, we consider a realistic center-pivot irrigation farm (i.e., the Al Jouf Arable farm) in Saudi Arabia, as illustrated in Fig.~\ref{fig_simscenarios}. The fully buried WUSN is used as the baseline since it co-locates sensing and wireless functions at the target depth, providing the most direct reference for evaluating the benefit of Terra-Fi's depth decoupling. The key simulation parameters are summarized in Table~\ref{ParaTab}.

We consider a large-scale agricultural monitoring area with a deployment radius of $5$~km, where $N$ UDs are randomly and uniformly distributed across the field. The target root zones are located at meter-scale depths, e.g., $1$~m and $2$~m, where in-situ measurements of VWC, soil temperature, conductivity, pH, and nutrient-related indicators are valuable for irrigation scheduling, fertilizer control, and crop-stress assessment. Fully buried WUSNs require the entire UD to be deployed at the root-zone depth, whereas \emph{Terra-Fi} keeps the wireless hub at $0.3$~m and extends only the sensing probes to the root-zone depth through power-and-signal feeders.

During each reporting period $T$, the UD first harvests energy from external power sources through WET during $T_w$, and then utilizes the harvested energy to sense and upload a $10$-byte packet to the cloud through an aboveground gateway located at a height of $10$~m during $T_t=T-T_w$~\footnote{WET is not considered for the fully buried WUSN baseline since direct WET to UDs at meter-scale depths is infeasible.}. In this study, LoRaWAN is adopted as the wireless communication technology due to its long-range coverage, low power consumption, low hardware cost, and subscription-free operation. Therefore, the system operates in the $433$~MHz ISM band, where all UDs are configured with a transmit power of $14$~dBm, an antenna gain of $2.15$~dBi, and a fixed LoRaWAN spreading factor (SF) without adaptive data rate. Meanwhile, $16$ uplink channels are used for uplink communication by employing a channel bandwidth of $125$~kHz and SF configurations ranging from SF7 to SF12. Moreover, we adopt the modified Friis model based on the mineralogy-based soil dielectric model and set the in-situ soil characteristics, e.g., clay percentage, to capture the UG2AG path loss~\cite{LinMag, LinLoRaWUSNs}. The aboveground propagation component of both the communication and WET links is assumed to be line-of-sight, with a path-loss exponent of $\eta=2$.
%Accordingly, the time-on-air (ToA) of a single $10$-byte uplink packet is~$T_{toa}=\{61.696, 113.152, 155.648, 370.688, 823.296, 1482.752\}$~ms for SF$7$--SF$12$, respectively.

We utilize the packet delivery ratio (PDR) to characterize LoRaWAN-based uplink reliability. Specifically, the PDR accounts for two components: (i) the probability that the received signal-to-noise ratio (SNR) exceeds the SF-specific SNR demodulation threshold; and (ii) the probability that the packet is successfully demodulated under co-channel interference by exploiting the capture effect, where the signal-to-interference ratio (SIR) of the target packet must exceed the capture threshold $\gamma=6$~dB. Furthermore, the energy per packet (EPP) quantifies the average energy required for a UD to successfully deliver one packet to the gateway, which is calculated as the ratio of the energy consumed per packet to the success probability. To accurately estimate the average operational lifetime across all UDs, we adopt the complete LoRaWAN Class A end-device energy consumption model described in~\cite{LinMag}. We assume that the shallow hub in \emph{Terra-Fi} can harvest RF energy through RF-WET with a harvested RF power of $0.3$~mW~\cite{UAVRFUWET}, and further consider a $3.3$~V operating voltage, a $114.15$~mA transmit current, a $3000$~mAh battery capacity, and an energy-conversion efficiency of $0.6$~\cite{LinMag}.

\begin{table}[!t]
\caption{Key Simulation Parameters} \label{ParaTab}
\centering
\begin{tabular}{m{0.3\textwidth}<{\raggedright} m{0.14\textwidth}<{\centering}}%

\toprule
\textbf{Parameters} & \textbf{Values} \\  
\hline
\multicolumn{2}{l}{\textbf{Deployment Environment}}\\ 
\hline
Deployment radius                  & $5$~km \\
Gateway height                       & $10$~m \\
Number of UDs   ($N$)                     & $10$k--$100$k \\
Depth of \emph{Terra-Fi} hub     & $0.3$~m \\
Depth of fully buried WUSNs      & $1$~m, $2$~m \\
VWC                                & $20\%$ \\
Clay content                      & $16.86\%$ \\
Frame payload                  & $10$ bytes\\
Reporting period ($T$)                  & $1800$~s \\
Traffic pattern                           & Periodic \\
\hline

\multicolumn{2}{l}{\textbf{Radio Configuration}}\\ 
\hline 
Carrier frequency             & $433$~MHz \\
Transmit power of UDs       & $14$~dBm \\
Antenna gains of UDs and gateway & $2.15$~dBi, $5$~dBi \\
Aboveground path-loss exponent ($\eta$)   & $2$ \\
\hline

\multicolumn{2}{l}{\textbf{LoRaWAN Configuration}}\\  
\hline
SF                                 & SF7--SF12 \\
Bandwidth                     & $125$~kHz \\
Frequency channels         & $16$ \\
SIR threshold ($\gamma)$           & $6$~dB \\
Noise power       & $-117$~dBm \\
\hline

\multicolumn{2}{l}{\textbf{Energy Model}}\\  
\hline
Operating voltage of UDs             & $3.3$~V \\
Transmit current              & $114.15$~mA   \\
Harvested RF power         & $0.3$~mW \\
Battery capacity                  & $3000$~mAh \\
Energy-conversion efficiency      & $0.6$ \\
\bottomrule
\end{tabular}
\end{table}

\subsection{Numerical Results and Discussion}
To evaluate the effects of SF adaptation and WET, we compare \emph{Terra-Fi} with fully buried WUSNs placed at the sensing depth. The following five strategies are considered:
\begin{itemize}
    \item \textbf{Fully buried WUSNs with SF12:} All UDs use fixed SF12 for data transmission.

    \item \textbf{Fully buried WUSNs with optimal SF:} All UDs are configured with the optimized SF that maximizes the average lifetime. 

    \item \textbf{\emph{Terra-Fi} with SF12:} The shallow hub uses fixed SF12 for wireless connectivity.

    \item \textbf{\emph{Terra-Fi} with optimal SF:} The optimal SF is selected for the shallow hub to maximize the average lifetime.

    \item \textbf{\emph{Terra-Fi} with optimal SF and RF-WET:} The shallow hub integrates SF optimization with RF-WET-based energy replenishment.
\end{itemize}

\begin{figure}[!t]
\centering
\includegraphics[width=3.1in]{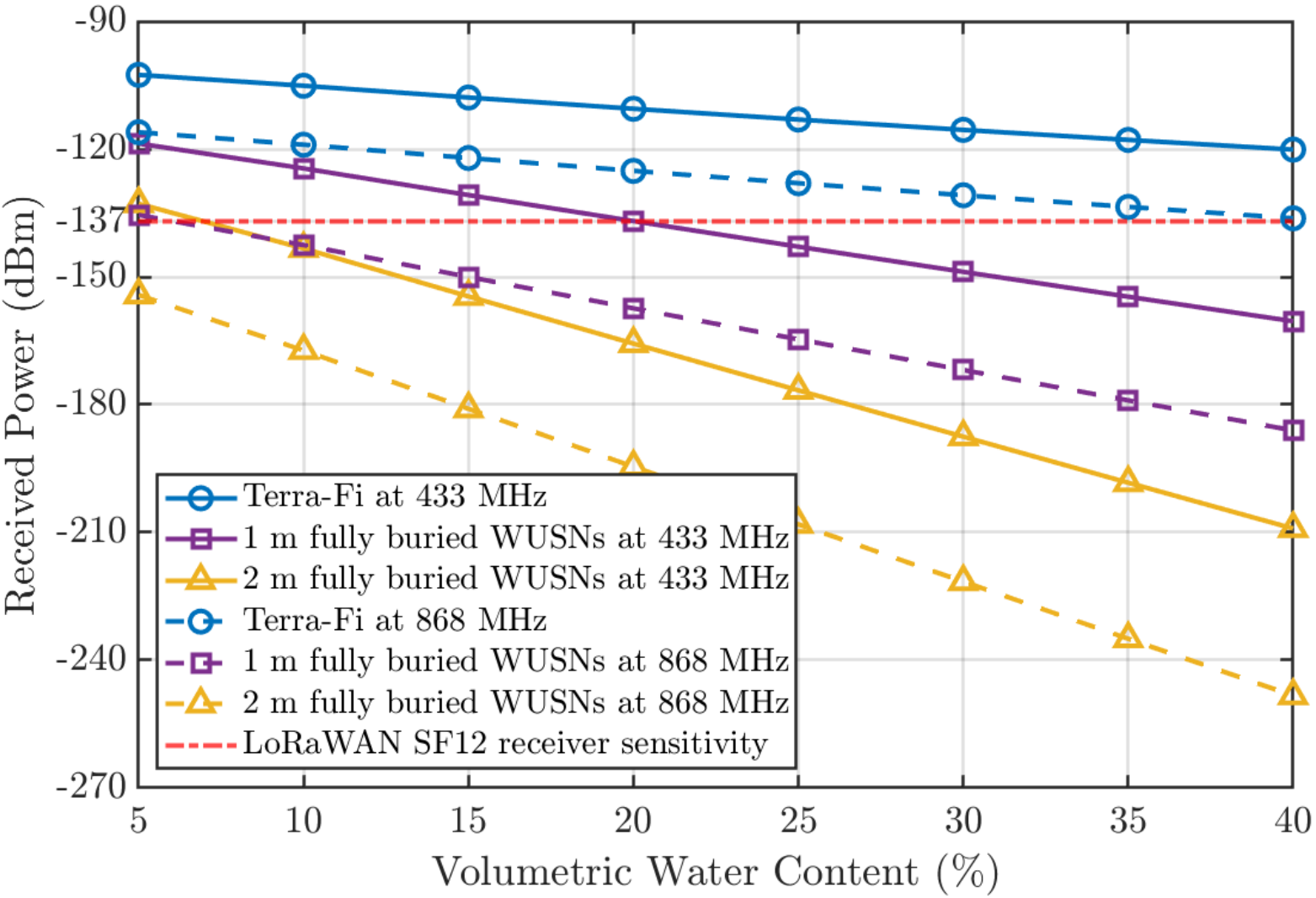}
\caption{Received signal power as a function of VWC at $433$~MHz and $868$~MHz for the proposed \emph{Terra-Fi} and fully buried WUSNs at depths of $1$~m and $2$~m. The bottom solid line depicts the receiver sensitivity threshold for LoRaWAN SF12.}
\label{PowerRes}
\end{figure}

First, we evaluate the feasibility of uplink packet transmission from a UD to an aboveground gateway. Fig.~\ref{PowerRes} illustrates the effects of VWC and frequency band on the received power under the proposed \emph{Terra-Fi} and fully buried WUSN architectures at depths of $1$~m and $2$~m. The LoRaWAN SF12 receiver sensitivity threshold is also shown to indicate the minimum received power required for successful packet reception in the absence of packet collisions. As expected, the received power decreases as the VWC increases because wetter soil introduces stronger electromagnetic absorption and thus larger underground attenuation. This effect becomes more pronounced as the burial depth increases. For instance, the received power of \emph{Terra-Fi} at both frequency bands remains above the SF12 sensitivity threshold across the entire VWC range, while the received power of the fully buried WUSN at $433$~MHz and depths of $1$~m and $2$~m decreases from $-119$~dBm and $-132$~dBm to $-160$~dBm and $-209$~dBm, respectively, as the VWC increases from $5\%$ to $40\%$. Moreover, $433$~MHz consistently outperforms $868$~MHz across all considered deployment cases since lower-frequency signals generally experience smaller dielectric loss in soil. However, even at $433$~MHz, fully buried WUSNs cannot maintain reliable UG2AG connectivity at large burial depths under high-VWC conditions. In contrast, the proposed \emph{Terra-Fi} architecture avoids this severe link-budget loss by placing the wireless hub at a shallow depth of $0.3$~m and extending only the sensing probes to the target root-zone depth.

\begin{figure}[!t]
\centering
\subfloat[]{
    \includegraphics[width=3.3in]{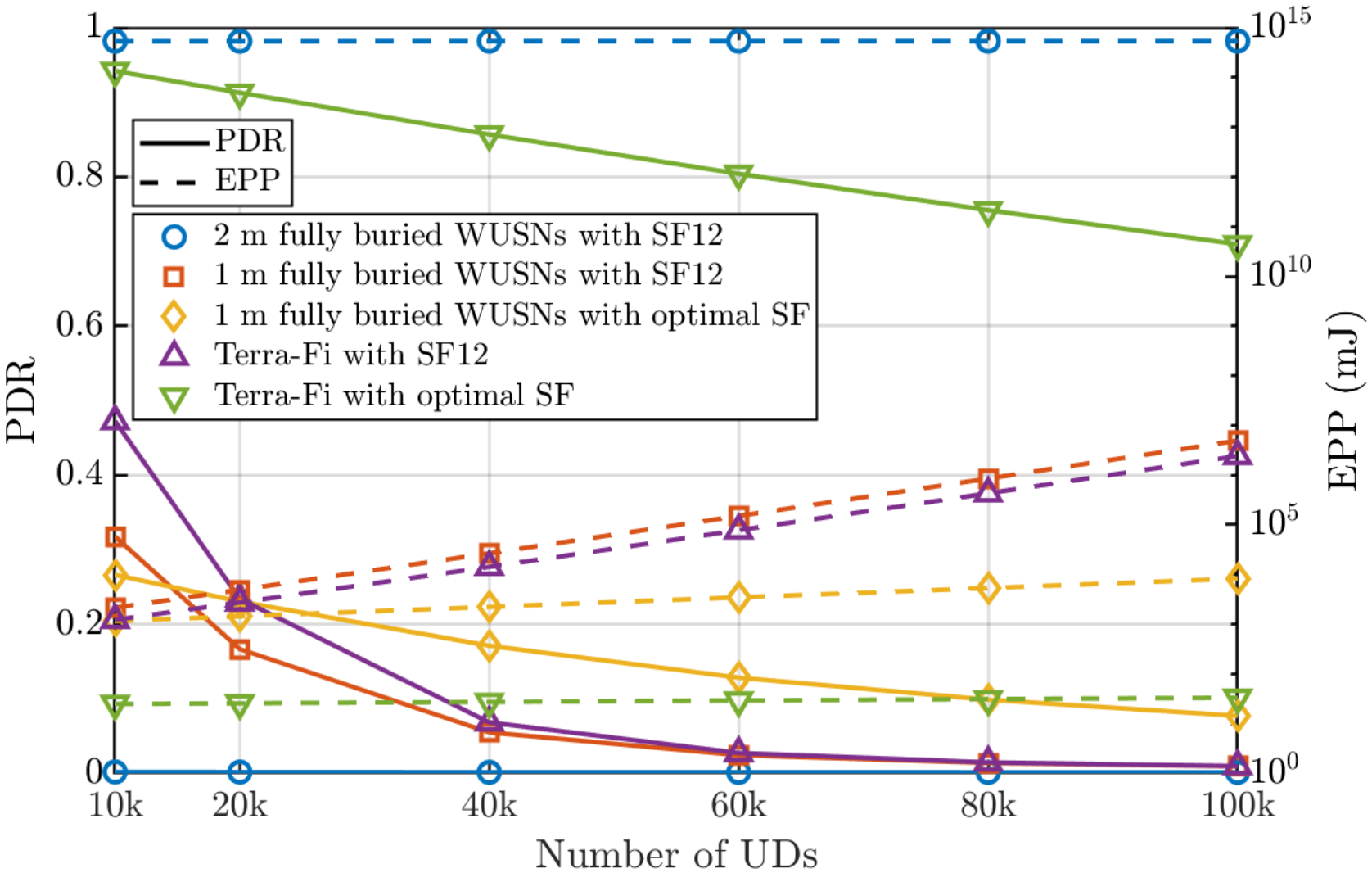}%
    \label{PDREPPRes}
}\hfil
\subfloat[]{
    \includegraphics[width=3.1in]{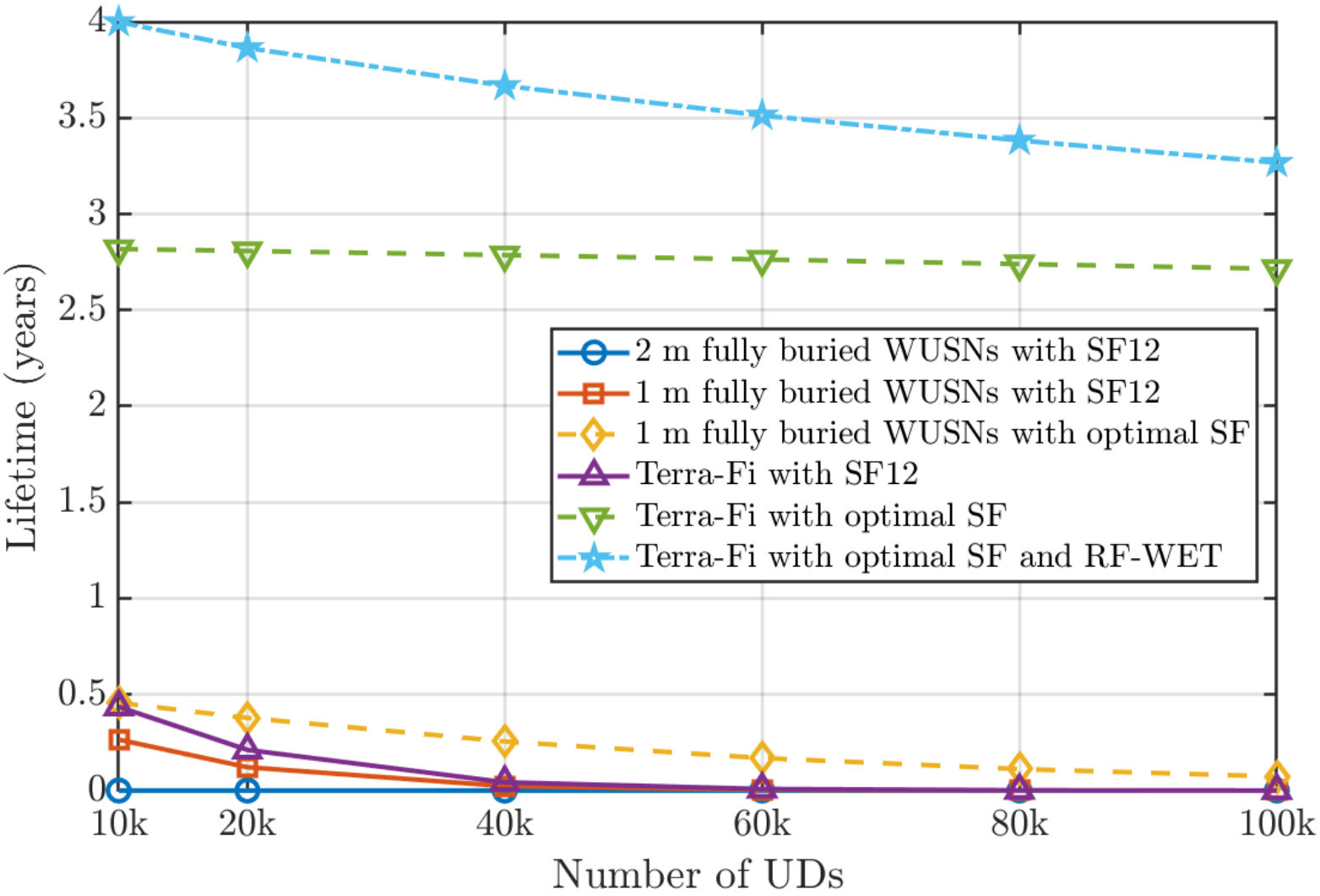}%
    \label{LifetimeRes}
}
\caption{(a) Average PDR and EPP across all UDs versus the number of UDs for the proposed \emph{Terra-Fi} architecture and fully buried WUSNs at depths of $1$~m and $2$~m under fixed and optimal SF configurations; (b) average lifetime across all UDs versus the number of UDs for fully buried WUSNs and \emph{Terra-Fi}, considering fixed and optimal SF configurations with and without RF-WET.}
\end{figure}

Then, we investigate the impact of network scale and SF configuration on the average PDR and EPP across UDs for the proposed \emph{Terra-Fi} and fully buried WUSN architectures. As shown in Fig.~\ref{PDREPPRes}, increasing the number of UDs raises the collision probability, thereby reducing PDR and increasing EPP, particularly for deeply buried UDs with limited link margin. For instance, the PDR of the fully buried WUSN at a depth of $2$~m is close to $0$, indicating that direct uplink transmission from a depth of $2$~m is infeasible even when using LoRaWAN SF12 featuring the highest link budget among the considered SF configurations. At a depth of $1$~m, SF12 achieves a PDR of $0.32$ for $10,000$ (i.e., $10$k) UDs, but its long time-on-air results in a high collision probability at larger network scales, reducing the PDR to $0.01$ and increasing the EPP to $4908.3$~J for $100$k UDs. In comparison, the \emph{Terra-Fi} shallow hub with SF12 deployed at a depth of $0.3$~m improves the PDR to $0.47$ for $10$k UDs, but remains collision-limited under dense deployments. By optimizing the SF, SF10 is selected for the fully buried WUSN at a depth of $1$~m across all network sizes, reducing the EPP to $8.24$~J at $100$k UDs. Furthermore, SF optimization selects SF7 for \emph{Terra-Fi}, enabling a PDR above $0.7$ and an EPP below $0.034$~J even at $100$k UDs. These results indicate that \emph{Terra-Fi} improves not only the UG2AG link budget but also the range of viable LoRaWAN transmission configurations in dense networks.

Finally, we analyze the impact of network scale, SF configuration, and RF-WET integration on the average operational lifetime across UDs under the five considered strategies. For \emph{Terra-Fi} with optimal SF and RF-WET, bounded numerical optimization is employed to determine the optimal WET duration for each SF configuration, followed by selecting the SF that maximizes the system lifetime. As depicted in Fig.~\ref{LifetimeRes}, the lifetime decreases as the number of UDs increases because denser deployments lead to a higher collision probability and larger EPP. The fully buried WUSN at a depth of $2$~m with SF12 remains non-operational across all network scales, while the lifetime of the fully buried WUSN at a depth of $1$~m with SF12 drops from $0.265$ years at $10$k UDs to nearly $0$ at $100$k UDs. With SF optimization, the fully buried WUSN at a depth of $1$~m selects SF10 and improves its lifetime to $0.46$ years at $10$k UDs and $0.07$ years at $100$k UDs. Meanwhile, \emph{Terra-Fi} selects SF7 due to its shallow hub deployment at a depth of $0.3$~m, maintaining a lifetime above $2.7$ years across all network scales. When RF-WET is further integrated with optimized time allocation, the lifetime of \emph{Terra-Fi} with optimal SF and RF-WET extends to $4$ years at $10$k UDs and remains $3.3$ years even at $100$k UDs. These results demonstrate that \emph{Terra-Fi} supports scalable long-term underground monitoring by combining shallow-depth wireless communication, appropriate SF configuration, and RF-WET-aided energy replenishment with optimized time allocation.

\section{Open Challenges and Research Directions}
The above quantitative analysis demonstrates the feasibility of \emph{Terra-Fi} for precision agriculture by decoupling the wireless operation depth from the target sensing depth through power-and-signal feeders. Note that \emph{Terra-Fi} can be generalized to other underground monitoring scenarios, such as pipeline leakage detection, subterranean environmental surveillance, and landslide early warning, by adapting the sensing probes and deployment configurations. Nevertheless, several practical challenges must be addressed before \emph{Terra-Fi} can be deployed in real underground monitoring systems.

\begin{itemize}
    \item \textbf{Feeder Electrical and Mechanical Reliability:} 
    The power-and-signal feeder is a critical component of \emph{Terra-Fi}, and its feasible length is jointly constrained by conductor resistance, probe power demand, voltage drop, power loss, and signal integrity. Long-term deployment further requires waterproof and corrosion-resistant conduits, sealed connectors, and mechanical protection against bending, soil compaction, pulling, and installation-induced stress. Therefore, it is essential to quantify these electrical and mechanical limits under representative feeder lengths and burial conditions.

    \item \textbf{Multi-Probe Interface and Interoperability:} 
    A single shallow hub may serve multiple probes with different supply voltages, sensing interfaces, and communication protocols. Modular probe interfaces should therefore support voltage regulation, signal conditioning, protocol conversion, individual addressing, power gating, and fault isolation. Such interfaces are particularly important for extending \emph{Terra-Fi} beyond application-specific sensor sets toward heterogeneous multi-depth sensing platforms.

    \item \textbf{Efficient Energy Management:} \emph{Terra-Fi} turns the shallow hub into a shared energy-service center that decides when to harvest energy, which probe or depth layer to power, how frequently to sample, and how much data to report. This calls for joint optimization of WET scheduling, stored-energy allocation, probe activation, depth-aware sampling, and transmission configuration under heterogeneous sensing demands, age-of-information requirements, and operational-lifetime constraints.

    % \item \textbf{Depth-Aware Data Fusion:} Multi-depth measurements should be converted into actionable indicators rather than treated as isolated sensor readings. Hence, it is worthwhile to develop depth-aware sensing frameworks that integrate multi-depth measurements with physics-based models, including soil-water transport, subsurface flow, structural response, and slope stability, while leveraging AI-assisted data-driven decision engines to translate underground observations into application-level decisions, such as irrigation scheduling, pipeline anomaly detection, contaminant tracking, and early geohazard warning.

    \item \textbf{Effective Deployment Planning and Installation:} \emph{Terra-Fi} introduces additional deployment variables, including hub depth, feeder length, probe-depth allocation, gateway density, and charger placement. These variables should be jointly optimized to balance sensing coverage, communication reliability, WET feasibility, installation cost, and maintenance accessibility. Furthermore, practical deployment can use small-borehole or direct-push methods to install the deep sensing probes and their feeders without full-depth excavation, while the hub remains at a shallow service depth.
\end{itemize}

\section{Conclusion}
To enable meter-scale underground monitoring, we proposed \emph{Terra-Fi}, a shallow-hub and deep-sensing architecture that decoupled the sensing depth from the wireless communication and energy-replenishment depth through power-and-signal feeders. After examining the depth limitations of underground wireless communication and WET, we presented the system architecture, operation flow, and key architectural tradeoffs. Through a realistic precision agriculture case study, our numerical results evidenced that \emph{Terra-Fi} achieved more reliable wireless communication and a longer operational lifetime than conventional fully buried WUSNs across different network scales, particularly under deep sensing zones and high-VWC conditions. Furthermore, the system performance was enhanced by jointly optimizing LoRaWAN SF configurations and the time allocation between WET and data transmission. Finally, we identified key open challenges to facilitate the practical deployment of this novel architecture.

\bibliographystyle{IEEEtran}
\bibliography{ref}

\section*{Biographies}
\footnotesize
\noindent \textbf{Kaiqiang Lin} [S'18, M'24] (kaiqiang.lin@kaust.edu.sa) is a Postdoctoral Research Fellow at the Computer, Electrical and Mathematical Sciences and Engineering Division (CEMSE), King Abdullah University of Science and Technology (KAUST). His current research interests include Internet of underground things, LoRaWAN, and sustainable underground monitoring.

\vspace{2mm}
\noindent \textbf{Mohamed-Slim Alouini} [S'94, M'98, SM'03, F'09] (slim.alouini@kaust.edu.sa) is the Al-Khawarizmi Distinguished Professor of CEMSE, the principal investigator of the Communication Theory Lab at KAUST. His current research interests include the modeling, design, and performance analysis of wireless communication systems.

\end{document}